# Marginal Gains or Meaningful Progress? Exploring Tech Tuber Narratives on Annual Smartphone Innovation

## Full Research Paper

### Chandima Wickramatunga

Department of Information Systems and Business Analytics
Deakin University
Burwood, Melbourne, Australia
Email: chandima.wickramatunga@deakin.edu.au

### Ruwan Nagahawatta

Sydney International School of Technology and Commerce
Melbourne, Australia
Email: ruwan.t@sistc.edu.au

### Anagi Gamachchi

Department of Information Systems and Business Analytics
Deakin University
Burwood, Melbourne, Australia
Email: a.gamachchi@deakin.edu.au

### Chintha Kaluarachchi

Department of Information Systems and Business Analytics
Deakin University
Burwood, Melbourne, Australia
Email: c.kaluarachchi@deakin.edu.au

## Abstract

Smartphone manufacturers continue to release new models annually, yet the pace of meaningful innovation has slowed, with most changes limited to incremental updates in design, performance, or software. This study examines whether such updates deliver tangible user benefits, as perceived by expert reviewers. Using a grounded theory approach, guided by Rogers' Diffusion of Innovation (DOI) framework, the research analyses reviewer discourse from 2021–2025 across three technology commentators. The analysis identifies three interrelated processes sustaining perceptions of innovation: innovation displacement, capability–utility divergence, and market complacency cycles. While some improvements are acknowledged, such as refined aesthetics or extended software support; they are seldom judged sufficient to justify annual releases. These findings highlight a growing disconnect between industry narratives of innovation and expert evaluations of value, raising questions about the strategic and environmental legitimacy of frequent upgrades. The study contributes to debates on responsible innovation, perceived value, and sustainable technology consumption.

# 1 Introduction

Smartphones were once seen as one of the most transformative innovations of the digital era. Between 2007 and 2016, rapid advances in touchscreens, app ecosystems, biometric authentication, mobile payments, and organic light-emitting diode (OLED) displays significantly reshaped digital life (Ohshima 2014). In recent years, however, innovation has slowed to incremental updates, mostly in processor speed, camera quality, or battery life (Kamal Hussain 2023). While these improvements remain valuable, they are no longer transformative. processing power now far exceeds the needs of most users, and mid-tier devices easily handle everyday tasks. Similarly, camera upgrades often deliver marginal benefits to typical users. The core smartphone experience has remained largely unchanged. As a result, the market is increasingly viewed as saturated and fatigued, with annual releases relying more on marketing than on meaningful breakthroughs (Brownlee 2024; Jerry 2025; Maini 2025). This perceived stagnation raises questions about the role and direction of innovation. Innovation, in this context, is not just desirable but essential to align product development with user needs and sustainability goals (Watz and Hallstedt 2022). Despite this plateau, manufacturers continue to promote each release as a significant leap forward. For example, Apple's iOS 18 was positioned as a major AI upgrade, yet many of the most publicised features were unavailable at launch (Apple 2024; Maini 2024). When later released, their performance fell short of expectations set during the keynote event. This gap between promotional claims and functional delivery reflects a broader trend: the symbolic framing of marginal updates as major innovation.

To explore how such framing is constructed and received, this study analyses the discourse of expert tech reviewers on platforms such as YouTube. These reviewers test devices hands-on, compare models across generations, and evaluate features from both technical and user perspectives. Their commentary offers a grounded view into how innovation is perceived by those with early and critical access to each release. To guide this interpretation, the study uses Diffusion of Innovation (DOI) theory (Rogers 1983; Rogers et al. 2014) not to study adoption, but as a sensitising lens to examine how new features are framed. The five DOI characteristics; relative advantage, compatibility, complexity, trialability, and observability; help reveal how perceived value is constructed or questioned in reviewer discourse (Rogers 1983). This approach allows the study to focus on perceived meaningfulness without making technical judgements.

Therefore, this study examines how the changes introduced in annual smartphone releases are perceived by expert reviewers, and whether they truly represent meaningful improvements. Specifically, it asks: *"How are annual smartphone releases perceived in terms of delivering meaningful improvements that translate into tangible and practical benefits for users?"*

To answer the research question, the study is structured as follows. It begins with a theoretical background, followed by the methodology used to analyse tech-tuber reviews. The discussion section presents the key findings, and the conclusion outlines the main insights, contributions, limitations, and suggestions for future research.

# 2 Theoretical Background

## 2.1 Diffusion of Innovation Theory

The DOI is a process-based theory for describing how, why, and how technology is adopted (Rogers 1983; Rogers et al. 2014). It describes the pattern of adoption, illustrates the process, and helps understand whether and how the adoption of technology will be successful (Rogers 1983; Rogers et al. 2019). DOI is also used to describe the process of technology diffusion as the process by which an innovation is transmitted across certain channels over time among participants of a social system (Rogers et al. 2014; Simões et al. 2020). In recent consumer markets, innovation is often linked with disruption, progress, and significant technological progress. However, in product sectors like smartphones, innovation has mostly become incremental; shown through small changes in design, processing speed, camera quality, or software features (Cecere et al. 2015; Jean 2017; Nagahawatta 2022; Prakash 2022). Despite these minor updates, new products are still promoted as innovation milestones. This situation raises important questions about how innovation is defined and recognised when there are no major technological breakthroughs.

Rather than viewing innovation as an inherent trait of technological artefacts, this study conceptualises it as a discursively constructed and socially mediated process influenced by marketing, media, and public discourse. To examine how marginal changes are legitimised as innovation, this section draws on DOI theory (Rogers 1983). While DOI is usually used to explain the spread and adoption of innovations, this study approaches the theory differently; focusing not on behavioural adoption but on how innovation is perceived, represented, and legitimised through discourse and performance.

Rogers (1983), theory of Diffusion of Innovation outlines five key attributes that influence how an innovation is perceived: relative advantage, compatibility, complexity, trialability, and observability. While these attributes are often used as predictors of adoption behaviour, they can also be viewed as discursive and symbolic constructs that help shape the framing of innovation; particularly when technological progress is limited. In saturated markets, these features rhetorically uphold the notion of progress, even when the material differences between product generations are minimal (Greenhalgh et al. 2004; Nagahawatta 2019; Prakash 2022). Relative advantage, defined as the perceived superiority of a new product over its predecessor, is central to this concept. In smartphone marketing, minor improvements, such as a slightly faster processor, a modestly improved camera, or incremental battery life gains; are often presented as breakthroughs (Kaluarachchi et al. 2020; Nagahawatta 2020; Schwartz 2022). Complexity relates to ease of use, and in marketing it is downplayed by presenting AI functions as seamless background processes (Jiang 2024; Nagahawatta 2025; Nama 2023). Trialability is the ability to experiment with innovations, now often simulated through launch events and influencer hands-on videos rather than direct experience. Observability concerns visibility of benefits, which is amplified by stylised online demonstrations that highlight visible change while masking limited everyday impact.

Taken together, DOI attributes function less as neutral descriptors than as symbolic tools that transform incremental updates into perceived innovation events.

## 2.2 Technology YouTubers and Expert Opinion

The term YouTubers refers to content creators, who record their videos on the most diverse subjects and manage their channels in the virtual environment hosted on the World Wide Web. They are the main responsible for contents published on the YouTube platform. Most YouTubers are young, and, in their channels, one can find a variety of videos on the most diverse themes, from tutorials to personal subjects, from video games to humour, music and a multitude of contents (Paganini et al. 2021).

Tech-tubers, YouTube creators reviewing consumer tech, play a unique role in the innovation debate. Often seen as promotional figures, their content increasingly reveals the limits of innovation, especially when updates are minor. Instead of emphasising new features, many serve as critical observers, highlighting the sameness of yearly cycles. Popular formats like "What's New?", "Is It Worth Upgrading?", and "Don't Buy This Yet" focus on comparing devices, with comments like "not a big upgrade" indicating that changes are more symbolic than functional (McQuarrie et al. 2013).

Visual techniques, such as side-by-side comparisons and split screens, demonstrate how little has changed, thereby challenging marketing claims (Wortmeier et al. 2024). The tone reflects fatigue, with creators and audiences becoming desensitised to novelty and questioning branding of minor improvements as innovation (Sachdeva 2020). Despite this scepticism, tech-tubers still operate within monetised ecosystems, but their commentary often exposes the inflation of innovation claims, making their reviews valuable for revealing minor updates.

## 3 Research Methodology

This research employs a qualitative research approach, utilising web content analysis to examine narratives to answer the research questions. First a literature review was conducted to understand the study context. According to Creswell and Creswell (2017), theories in qualitative research can serve two main purposes: (1) to generate a theory as the outcome of the study, or (2) to employ an existing theory as a lens to inform the research process, shaping what to observe and which questions to ask (Creswell & Creswell, 2017). This study adopts the latter approach. Rather than testing a theory, it uses Rogers' innovation characteristics as a sensitising framework to inform the analysis and guide interpretation of emergent themes, without constraining the inductive coding process.

To investigate the nature of meaningful innovation in flagship smartphones, this study focuses on Apple and Samsung as the two dominant premium manufacturers, collectively holding over 49% of global smartphone market share (StatCounter 2025) and setting industry benchmarks. This study draws on content produced by three prominent YouTube technology reviewers: MKBHD, Mrwhosetheboss, and JerryRigEverything, spanning five years from 2021-2025 (inclusive). The five-year timeframe captures multiple product generations, enabling identification of longitudinal innovation patterns across 31 analysed video transcripts. MKBHD (coded as reviewer 01), with over 20.2 million subscribers, is known for his in-depth smartphone reviews and consistent testing protocols. His content provides comprehensive evaluations of flagship devices, often employing repeatable and objective criteria. Mrwhosetheboss (coded as reviewer 02), followed by over 21.4 million viewers, focuses on comparative smartphone reviews and annual innovation roundups. His analyses highlight inter-brand competition and year-over-year advancements. JerryRigEverything (coded as reviewer 03), whose channel has more than 9.48 million subscribers, specialises in durability tests and teardowns. His systematic disassembly of devices reveals underlying design changes and engineering decisions not visible through surface-level reviews. These reviewers were selected due to their consistent access to early-release flagship devices, rigorous and transparent testing methodologies, and influence within the broader technology community. Their large audiences and longstanding reputations lend credibility to their analyses. Importantly, their combined expertise offers triangulated coverage of key innovation dimensions: performance and usability (Reviewer 01), comparative market positioning (Reviewer 02), and hardware construction (Reviewer 03). This approach reduces the risk of single-source bias and ensures that multiple facets of smartphone innovation are addressed.

### 3.1 Coding Structure and Data Analysis

This study employed a grounded theory methodology, drawing on the work of Glaser and Strauss (1998) and further guided by recommendations from (Williams and Moser 2019). To ensure methodological rigor, each video was reviewed multiple times to support accurate interpretation and comprehensive data capture. The content was systematically analysed using NVivo qualitative analysis software, following the stages of open, axial, and selective coding.

Open coding was first used to inductively identify initial references to innovation and developments within the expert reviews, capturing granular mentions related to hardware design, functional enhancements, feature usability, and user experience. These open codes were then organised through axial coding, which grouped them into conceptually coherent categories based on patterns in how reviewers described the implementation, practicality, and value of new features (e.g., "Feature fatigue and interface clutter" or "Delayed feature rollouts") (Glaser and Strauss 1998). Thereafter, to deepen theoretical interpretation and ensure analytical alignment with the study's focus on innovation perception, the five attributes of Rogers' Diffusion of Innovation theory; relative advantage, compatibility, complexity, trialability, and observability, were used as sensitising concepts. These five characteristics served as the final thematic categories, under which the axial codes were grouped. This approach allowed the study to retain grounded, data-driven insight while simultaneously situating the findings within a well-established theoretical framework for understanding how innovations are perceived and adopted. This step enabled differentiation between genuinely novel advancements and more incremental or cosmetic changes. This use of Rogers' DOI theory does not replace the inductive logic of the study but rather offers a conceptual frame to organise and interpret emerging themes within a broader theoretical context. By sensitising the axial codes through the lens of DOI, the study retains interpretive openness while grounding its insights in an established model of innovation perception (Bowen 2006).

To ensure reliability, the coding process was conducted independently by two researchers. A subset of video transcripts was double-coded, and coding alignment was discussed in collaborative sessions with the research team. Intercoder reliability was evaluated using Krippendorff's alpha, with agreement exceeding the 75% threshold recommended by Krippendorff (2018), ensuring consistency and trustworthiness in the interpretation of qualitative data.

Table 1 presents the sample coding structure used in this study. Open codes derived from reviewer discourse were first grouped into axial codes based on conceptual similarity. These axial codes were then sensitised through Rogers' (1983) innovation attributes and organised under five categories as relative advantage, compatibility, complexity, trialability, and observability, which together form the study's final thematic structure.

| **Verbatim Quote with Open Code** | **Sample Axial Codes** | **Selective Code** |
|---|---|---|
| "Camera rings on the outside are just a little more pronounced, fancy, you know. But none of this really makes a huge difference to usability *(No change to usability)*" – Reviewer 01 | Marketing minor changes as big | Relative Advantage |
| "And of course, I'm always impressed by Apple's ability to run an hour and a half event over what essentially boils down to slapping a new number on the box *(Big marketing events for tiny updates)*." - Reviewer 03 | Big marketing events for small changes | |
| "But the thing is, I can't find a way to actually use that power in my day to day life. Like I said, The phone does just feel faster than last year, but that's just the software. That's not because of the chip *(Marginal performance gains not noticeable in everyday use).*" – Reviewer 02 | Changes not needed | |

*Table 1: Sample Coding Structure*

## 4 Interpretation and Discussion

This study employed Rogers (1983) DOI framework as a sensitising lens during axial coding, allowing the emergent categories to be organised around five perceived innovation attributes: Relative Advantage, Compatibility, Observability, Complexity, and Trialability. Across 89 coded reviewer statements from high-profile technology commentators, these categories collectively portray an industry where the annual flagship release cycle is sustained more by marketing spectacle than by substantive technological progress. The following subsections present each selective code, integrating multiple axial codes and supporting quotes to illustrate the thematic patterns that emerged from this comprehensive analysis.

### 4.1 Relative Advantage

Rogers defines relative advantage as the degree to which an innovation is seen as better than what it replaces. In practice, reviewers repeatedly challenged whether flagship models genuinely delivered meaningful improvements, with this dimension emerging as the most prominent in the analysis, comprising 64% of all coded statements. This overwhelming focus suggests that the erosion of genuine relative advantage represents the primary mechanism through which industry stagnation manifests.

The most pervasive pattern involved marketing minor visual changes as major improvements, where manufacturers transform cosmetic modifications into perceived breakthroughs. Reviewer 03 mocked Apple's "ability to run an hour and a half event over what essentially boils down to slapping a new number on the box", while Reviewer 02 likened the iPhone design cycle to "rotating between different options to keep whatever's new feeling different", rather than pursuing genuine advancement. This theatrical approach extends to colour marketing, where Reviewer 02 observed manufacturers "tried to make them interesting with the names like the hero colour is called titanium silver, blue… Let's call a spade a spade. It's gray". Even camera modifications follow this pattern, with Reviewer 01 noting that "camera rings on the outside are just a little more pronounced, fancy, you know. But none of this really makes a huge difference to usability".

Perhaps more concerning is the emergence of performance improvements users cannot access. Hardware gains often translated into capabilities that remained largely theoretical in everyday use. Reviewer 02 admitted, "I can't find a way to actually use that power in my day-to-day life" despite significant chip improvements, while Reviewer 01 explained that "most mobile games are developed to run on $100 phones, so finding ones that actually benefit from this power is becoming quite a niche endeavor". Reviewer 02 mentions that this creates a fundamental disconnect where "the average person is going to notice compared to the S 24 Ultra, because most of the bottlenecks come down to software, not hardware". Even traditionally important metrics like battery life showed stagnation, with Reviewer 01 reporting, "battery life about the same as last year, to be honest, like, not noticeably better, not noticeably worse" despite processor efficiency claims.

The analysis revealed a troubling pattern of feature removal and product regression disguised as streamlining. Samsung's elimination of Bluetooth S Pen functions exemplified this trend, justified, as Reviewer 01 observed, because "less than 1% of people ever actually activated that feature". Yet this utilitarian approach contradicts flagship positioning, as Reviewer 03 noted: "after seeing the internals and the externals this year's S25 ultra just seems kind of like a downgrade from last year's S24 Ultra". Design constraints further compound regression, with manufacturers "removing features for aesthetics" by "taking their top end Galaxy S 25 Ultra and chopping things up" to achieve thinness.

These limitations create poor value propositions despite quality, where technical competence fails to justify pricing. Reviewer 02 highlighted the disconnect: "you can get a Galaxy S 24 Ultra brand spanking new for $800 whereas this is a $1,300 phone, it's like 60% more expensive" for marginal improvements. Reviewers frequently concluded devices were "not special enough to justify the cost" despite acknowledging build quality. This pricing disconnect extends to repair costs, with reviewer 03 questioning why Apple charges "$379 to replace a screen when it's so easy to replace", suggesting manufacturers leverage market position rather than offer proportional value.

Incomplete AI implementation emerged as the industry's primary strategy for maintaining perceived advantage, yet execution remained problematic. Features promised at launch often arrived later or underperformed, with Reviewer 02 noting "Apple intelligence… isn't actually coming till later in the year, which kind of makes this feel like a Google Pixel event where the product comes first and then the feature later". When AI features did function, reviewer 02 found they're "always 80% what you're looking for, but they're way less likely to be the 100%", while basic functions were rebranded, as he observed: "the daily briefs… It's barely AI".

This occurs within a context of market complacency reducing innovation, where established positions enable conservative strategies. Reviewer 01 observed that Samsung "established themselves as the go to safe option for so many people. So now that they have, of course, they're playing it safe", particularly because "in places like the US, where they're huge… they just actually don't have pressure to compete directly against the Vivo's and Xiaomi's". The result is what he described as moving from "why would you buy anything else to this year saying you could buy anything else", indicating competitive differentiation has largely disappeared.

The analysis also revealed regulatory compliance replacing innovation leadership, where legal requirements drive change rather than technological ambition. Reviewer 03 observed that Apple "might not be as Ultra generous as we thought, and they're only providing repair guides, cause soon they'll legally have to either way". Environmental messaging similarly lacks authenticity, with manufacturers talking "a big game about recycling and saving the planet, but the biggest thing they could do for Earth by far is to just take a gap year" rather than pushing annual releases.

Across the industry, an innovation plateau has emerged where systemic stagnation affects all major manufacturers. This manifests through marginal improvements; camera megapixel increases and marginal improvement in charging speeds, that generate marketing emphasis but deliver negligible and real practical impact. The plateau creates conditions where consumers extend device lifecycles despite marketing campaigns, while reviewers increasingly question annual upgrade necessity, indicating traditional competitive advantage mechanisms are losing effectiveness.

### 4.2 Compatibility

Compatibility refers to how well an innovation aligns with users' existing needs and values. The analysis revealed fundamental misalignments between manufacturer priorities and user preferences, creating systematic compatibility failures that undermine adoption motivation.

The most evident compatibility failure involved fundamental misalignment between flagship development priorities and user preferences. The pursuit of ultra-thin devices exemplified this disconnect, with reviewer 02 observing that "nobody asked for this, like these ultra-thin phones look cool… but what we really want is thicker phones that have less camera bump, but way more battery." This mismatch represents a systematic prioritisation of form over function, where manufacturers pursue design aesthetics that actively contradict expressed user needs for improved battery life and reduced camera protrusion.

Professional-oriented features increasingly dominated devices designed for mass markets, creating compatibility gaps for average users. Reviewer 02 noted that "a lot of the extra features now, while impressive, are tailored to actual industry professionals, as opposed to the casual user who just wants the best point and shoot camera". This resulted in features that were "really, really good feature, but for like, 2% of iPhone users", creating complexity without corresponding value for the majority of users. Video features particularly suffered from this professional focus, becoming "aimed at professionals, not average users" while casual users struggled with basic functionality.

The persistence of established market position enabling safe choices became evident in how dominant manufacturers approached innovation. Reviewer 01 observed that companies can "play safe because they can" when they've "established themselves as the go to safe option". This manifested in reactive rather than proactive development, where changes occurred because manufacturers were "making the legally required modifications to better comply with the new EU regulations" (Reviewer 03), rather than anticipating user needs. The comfortable market position enabled companies to pursue internal priorities while maintaining compatibility rhetoric.

Software fragmentation and regional inconsistency further compromised compatibility by creating artificial scarcity and confusion. Reviewer 01 observed that "most of the new stuff on this phone is this software and the AI stuff, I didn't really see a whole lot of reason why most of it isn't going to also be on the S 24 in a couple weeks", suggesting that technical limitations were not driving exclusivity. Reviewer 02 highlighted how this creates marketing confusion: "Apple intelligence… isn't exclusive to these phones, which isn't in itself a problem… but it just makes the launch event feel a little misleading", when core features work on older devices. Regional rollouts compound this issue, with "limited regional rollout of headline features" creating geographic compatibility gaps that undermine the universal appeal manufacturers claim.

## 4.3  Observability

Observability captures the extent to which the benefits of an innovation are visible to others. The analysis revealed a fundamental observability crisis where technological progress has become increasingly imperceptible to users, undermining the traditional mechanism through which innovations demonstrate their value.

The most striking pattern involved visual similarity to previous models that made differentiating between generations nearly impossible. Reviewer 01 described the S25 Ultra as looking "a lot like last year's $1,299 S 24 Ultra", while reviewer 03 noted the iPhone's "same titanium exoskeleton design that we saw last year". This similarity extended beyond surface aesthetics to fundamental design decisions, leading reviewer 01 to observe "this is basically a Galaxy S 24" rather than representing meaningful evolution. The visual continuity became so pronounced that "past models still viable alternatives" to current flagships, undermining traditional upgrade motivations.

Although technical upgrades were present, they often translated into performance gains that users could not see or feel. Despite significant hardware advances, the year-to-year experience remained largely unchanged, with battery technology in particular showing little progress despite repeated claims of improvement. Display technology illustrated this invisibility crisis most clearly: reviewers pointed to only "minimal changes to screen brightness and refresh rates" and the absence of any "major leap in display technology," even as marketing continued to highlight screen advancements. Even when improvements theoretically existed, they became invisible through other factors, such as screens that "seem to be consistently slightly dimmer" despite brightness claims (Reviewer 02).

The observability challenge was compounded by marketing confusion about new versus old features that blurred the line between genuine innovation and repackaging. Reviewer 02 described launch presentations as "telling you about some really cool stuff that the phone does, but then you slowly realising halfway through that the things they telling you were already features on last year's iPhone 15". This confusion arose from features rolled out in software updates but branded as hardware-linked, while older models gain same features via updates, creating situations where software parity across models undermines need to upgrade despite artificial marketing distinctions.

Marketing theatre and launch spectacles increasingly served as substitutes for genuine innovation, constructing a sense of observability where little real progress could be shown. These events evolved into elaborate productions aimed more at shaping perception than at demonstrating substantive advancements. Reviewer 03 mocked Apple's "ability to run an hour and a half event over what essentially boils down to slapping a new number on the box", while reviewer 01 noted "a whole lot of talk, a whole long section of the keynote that where they constantly talked about Apple intelligence" despite limited actual functionality. Reviewer 02 observed that these events featured "the same marketing narrative each year with minor tweaks" and "recycled talking points from previous launches", suggesting that visibility of innovation was being constructed through presentation rather than substance. The spectacle often relied on "sales pitch relies on selective comparisons" that highlighted metrics favourable to new devices while obscuring areas of stagnation.

## 4.4  Complexity

Complexity relates to the perceived difficulty of understanding or using an innovation. The analysis revealed how smartphone advancement has increasingly favoured technical sophistication over user accessibility, creating products that are simultaneously more capable and less user-friendly.

A significant trend involved professional features for the mass market that exceeded typical user needs and abilities. Arun observed that "a lot of the extra features now, while impressive, are tailored to actual industry professionals, as opposed to the casual user who just wants the best point and shoot camera", creating complexity hierarchies that served marketing more than users. This manifested in "Pro" branding that often lacked "matching benefits" for average users, where reviewer 01 noted you should "buy the iPhone 16 pro if you really want a better phone in cameras or display. That's really the two places where the pro shines", suggesting limited differentiation despite complexity additions. The result was "feature exclusivity tied to model rather than necessity" where reviewer 02 observed that Apple would "save all the really cutting-edge stuff... to release on the pro phones first", creating artificial complexity barriers.

Physical design constraints forced trade-offs that heightened user complexity while diminishing functionality. The drive for extreme thinness epitomised this dilemma by reviewers claiming that manufacturers reduced the thickness by taking their top end Galaxy S25 Ultra and chopping things up, rather than pursuing genuine engineering solutions. As a result, users encountered feature removals justified by aesthetics instead of benefiting from advances in materials science or innovative design approaches.

Unfinished AI functionality emerged as the most significant complexity challenge, with advanced features heavily promoted in marketing yet delivered in incomplete and unreliable forms. Reviewer 02 noted that cross-app communication "needs a couple of years in the oven" despite being presented as ready to use. This led to user experiences dependent on "post-launch software patches to fix day-one issues" (Reviewer 01) and devices burdened by bloatware that undermines out-of-box experience, where basic reliability coexisted uneasily with sophisticated but unstable AI capabilities.

The complexity challenge was amplified by technical complexity without user benefit, where specifications became ends in themselves rather than means to improved experience. Reviewer 03 detailed explanation of sapphire crystal thermal conductivity in camera buttons exemplified this pattern: extensive technical sophistication that users couldn't perceive or benefit from in daily use. Marketing increasingly featured "over-promotion of lab test results over lived experience" and "press focusing on headline specs over real-world use", where technical achievements failed to translate into practical advantages while adding conceptual complexity.

### 4.5 Trialability

Trialability denotes the extent to which an innovation can be experimented with before adoption. The analysis revealed how traditional trial mechanisms have been systematically undermined by industry practices that prioritise marketing cycles over user evaluation opportunities.

The most problematic pattern involved delayed feature availability that forced users to commit to devices based on promises rather than demonstrated capabilities. Reviewer 02 described how "Apple intelligence... isn't actually coming till later in the year, which kind of makes this feel like a Google Pixel event where the product comes first and then the feature later". This strategy of delay fundamentally reshaped the trial process, forcing users to commit to devices on the basis of promised potential rather than present functionality. AI features were especially affected, with "AI availability delayed after launch hype" leaving core marketing promises unfulfilled at the very moment when evaluation mattered most.

A concerning development involves manufacturers increasingly removing features on the basis of low usage data, even when those features offered value to specific user groups. As reviewer 01 noted, Samsung's decision to eliminate Bluetooth functions from the S Pen exemplifies this trend and reflects a broader industry tendency to streamline products for mass-market averages. While such removals may appear efficient, they diminish trialability by reducing opportunities for users to experiment with and discover novel capabilities, ultimately narrowing the scope of innovation.

Growing recognition emerged that extended development cycles needed to replace annual releases that had become incompatible with meaningful innovation development. Reviewer 03 suggested that "smartphone Marketing would be a heck of a lot easier if we went back to the biannual upgrade schedule. Can you imagine the hype of a new Samsung phone after two whole years of cooking in silence?", indicating that current cycles prevented adequate development time for trial-worthy innovations. This timing pressure created situations where reviewers suggesting skipping generations because consecutive years offered insufficient differentiation to justify upgrade costs, and where "annual cycle causes fatigue among buyers" who couldn't meaningfully evaluate marginal differences between rapid releases.

The trialability crisis was further compounded by the complexity of evaluating AI features that promised transformative capabilities but delivered inconsistent results. Users faced the challenge of trial-testing capabilities that were simultaneously sophisticated and unreliable, creating evaluation difficulties that traditional trial mechanisms couldn't address effectively.

In summary, the sources paint a picture of smartphone innovation having shifted from groundbreaking breakthroughs to a pattern of minor refinements, reliance on software promises, and even some feature removals, indicating a period of stagnation in truly revolutionary advancements where traditional innovation adoption mechanisms have been systematically undermined.

# 5 The Mechanics of Innovation Stagnation through Reviewer Discourse

In the flagship smartphone industry, annual product cycles have shifted innovation from technological breakthroughs to marketing-led perception strategies. Grounded Theory analysis of reviewer discourse shows how firms increasingly frame cosmetic or incremental changes as major advances (Relative Advantage), while neglecting user alignment (Compatibility). New features are often imperceptible in daily use (Observability), professional-grade tools or unfinished AI raise Complexity, and delayed rollouts or rigid annual cycles restrict Trialability, reinforcing consumer fatigue.

This analysis, based on 89 open codes clustered into 23 axial categories and sensitised through (Rogers 1983) five innovation attributes, demonstrates that annual releases introduce some new elements but present them as significant mainly through strategic marketing. Reviewers frequently noted that launch events had become elaborate productions amounting to little more than rebranding, underscoring how presentation has come to supersede substance.

To move from descriptive coding toward theoretical explanation, the study followed Eisenhardt (1989) and Eisenhardt and Graebner (2007) principles of theory building from qualitative data. Reviewer statements were first broken down into open codes that captured concrete observations and critiques. These were clustered into axial categories, which Rogers' five attributes were used to sensitise and organise. Through iterative comparison between coded data, DOI constructs, and emerging patterns, higher-level selective codes were abstracted. This process enabled the identification of three mechanisms that extend beyond Rogers' attributes to explain how innovation stagnation is actively sustained in the flagship smartphone industry.

## 5.1 Innovation Displacement

*Proposition 1: When superficial, highly visible changes are marketed as breakthroughs, they displace substantive innovation and sustain only the perception of progress.*

Cosmetic modifications such as colour variations, bezel adjustments, or material substitutions are presented as major advances, reinforcing observability and relative advantage without meaningful improvement. This redirects resources toward features that photograph well and generate launch spectacle, rather than toward technological development with lasting user value.

## 5.2 Capability–Utility Divergence

*Proposition 2: When technical specifications improve without corresponding user benefits, a divergence emerges between capability and utility that erodes innovation value.*

Benchmark gains in processor speed or memory often remain underutilised, as most apps and games are designed for lower-end devices. This widens the gap between device potential and everyday experience, reducing compatibility with user needs and increasing complexity while enabling firms to market technical capability as innovation.

## 5.3 Market Complacency Cycles

*Proposition 3: When firms rely on entrenched market positions and predictable release schedules, innovation becomes conservative, reinforcing cycles of complacency and weakening trialability.*

Dominant firms leverage their "safe option" status to release minimal annual upgrades, while flagship features quickly migrate to older models, eroding incentives to upgrade. This pattern sustains incrementalism and reduces differentiation, shifting the industry from genuine innovation leadership to routine, low-risk product refreshes.

## 5.4 Theoretical Integration

*Proposition 4: Innovation stagnation in mature technology markets is sustained through the interaction of displacement, divergence, and complacency, which collectively substitute manufactured differentiation for genuine progress.*

Together, the three mechanisms form a self-reinforcing system. Innovation displacement maintains upgrade momentum through visible but superficial changes. Capability–utility divergence enables continued technical marketing while masking limited real-world benefits. Market complacency cycles reduce competitive pressure for breakthrough innovation, while extended device lifecycles and reviewer recommendations to skip generations paradoxically reinforce manufacturer reliance on incrementalism.

This dynamic produces a sustainability paradox. Consumers rationally extend replacement cycles in response to minimal improvements, incidentally, reducing e-waste, while manufacturers persist with resource-intensive annual releases that serve marketing spectacle rather than user needs. In a functioning market, stagnation would push firms either toward genuine innovation or toward less frequent release schedules, yet these mechanisms allow wasteful production to continue.

This analysis shows how mature technology markets strategically manipulate Rogers' adoption attributes. Rather than naturally driving diffusion, relative advantage, compatibility, observability, complexity, and trialability are repurposed to maintain perception management. Extending innovation theory therefore requires recognising how adoption mechanisms can be strategically deployed to reinforce market rhythm and competitive positioning, even in the absence of substantive progress.

## 6　Conclusion, Contributions and Future Research Directions

This study examined how innovation in flagship smartphones is represented in expert reviews and whether it is perceived as meaningful enough to justify annual release cycles. Using grounded theory analysis of 89 open codes, consolidated into 23 axial codes, and organised through Rogers (1983) Diffusion of Innovation framework, the findings reveal recurring dissatisfaction with the pace, depth, and user relevance of change in these devices. While reviewers occasionally acknowledged incremental refinements such as marginal design adjustments or ecosystem consistency, they more often described three mechanisms of stagnation: innovation displacement (superficial changes as visible markers of progress), capability–utility divergence (technical specifications outpacing everyday usefulness), and market complacency cycles (entrenched positions reducing competitive urgency).

Across the dataset, these processes suggest that many annual "innovations" are perceived less as genuine advancements and more as strategies to sustain the appearance of novelty. When marketed breakthroughs fail to deliver observable, compatible, or advantageous improvements, a disconnect emerges between consumer expectations and user experience. In this dynamic, tech reviewers act not merely as commentators but as critical intermediaries who legitimise or challenge manufacturer narratives. By foregrounding the gap between release claims and perceived value, this study highlights tensions between commercial product cycles and the delivery of user-centred progress and underscores the potential value of longer development timelines (e.g., bi-annual release models).

Theoretically, this study extends DOI by applying it not only to adoption but also to perceived innovation stagnation, generating a grounded model of the flagship smartphone market's innovation deficit. Practically, it challenges the rationale for annual upgrades by demonstrating the mismatch between manufacturer narratives and expert evaluations, with implications for product development, marketing strategy, and communication practices. Methodologically, it illustrates how expert tech reviews provide a robust qualitative data source for grounded theory research in innovation and information systems.

These findings must be considered in light of certain limitations. Influencer content may still be shaped by sponsorships, algorithmic incentives, or brand affiliations. The focus on flagship devices from major Western brands limits generalisability to mid-range markets or non-Western contexts. Finally, the analysis captures expert perceptions but not consumer decision-making, leaving open questions about the influence of these evaluations on actual purchasing behaviour.

Future research could extend this work by incorporating user-generated discourse to explore wider audience engagement with innovation narratives. Studies of mid-range devices, emerging markets, or longitudinal user experiences may yield alternative perspectives. Linking reviewer perceptions with environmental disclosures and independent sustainability metrics could also test whether claims of technological progress align with measurable sustainability outcomes.

# 7　References

# Appendix

# Copyright